\documentclass[12pt]{iopart}

\usepackage{graphicx}
\begin{document}

\title{Excitation of positronium from 2S to 20P state}

\author{M. W. Heiss, G. Wichmann, B. Radics and P. Crivelli}

\address{Institute for Particle Physics and Astrophysics, ETH Zurich, Switzerland}
\ead{crivelli@phys.ethz.ch}
\vspace{10pt}
\begin{indented}
\item[]September 2018
\end{indented}

\begin{abstract}
We report the  observation of positronium excitation from the 2S to the 20P state. 
The Rydberg positronium atoms fly a distance of 40 mm before being field ionized and detected in a micro-channel plate. 
The time of flight can thus be measured and the velocity distribution of the atoms excited in the 2S state is reconstructed. 
This is used as an input to the model of the  line-shape in order to properly take into account the second order Doppler shift which is the main systematic uncertainty in the 1S-2S measurements of positronium.
\end{abstract}

%
%
%
%
%

\section{Introduction}

Positronium (Ps) is one of the simplest bound states governed by quantum electrodynamics (QED) \cite{SavelyPhysRep}.
Being a purely leptonic hydrogen-like atom, it can thus be used to perform precise test of bound state QED by measuring its properties, such as the lifetime \cite{Vallery:2003iz,Kataoka:2008hj,Adkins:2000fg,Adkins:2015jia}, the hyperfine splitting \cite{MillsHFS,RitterHFS,Ishida:2013waa,HFSTheoryBaker,HFSTheoryAdkins,HFSTheoryEides} and the transition frequencies via precision laser spectroscopy \cite{Fee:1993zz, Theory1S2SPachucki,Theory1S2SMelnikov}. 
In the last few years, tremendous advances in positronium laser physics have been achieved (see \cite{CassidyReview2018} for a recent review on this topic). This is mainly due to the advent of buffer gas traps \cite{DensePs,SurkoReview2015} that allow to create high Ps densities which can be probed with pulsed lasers. In fact, in a buffer gas trap positrons can be accumulated and ejected in intense pulses with a repetition rate of a few Hz comparable with standard pulsed laser systems  \cite{CassidyPRA2010, Aegis2018}. 
Moreover, Ps can be used to test fundamental symmetries \cite{asai2010,Moskal2016}, search for new physics beyond the Standard Model \cite{Badertscher:2006fm,Vigo:2018xzc} and efforts are ongoing to test its gravitational properties \cite{CassidyWAG2013,CrivelliQM2015,MillsRyd2018,QuPlus}. 
Furthermore, the application of positrons and positronium in characterization of materials with advanced functionalities on the nanoscopic scale is steadily growing \cite{Reviews}.

Being the lightest atom and due to its instability because of self-annihilation (the long lived triplet state has a lifetime of about 142 ns), Ps is a challenge for precision spectroscopy. In fact, at room temperature Ps has a velocity of approximately $10^5$ m/s. Cooling this atom via collisions with cold surfaces is very inefficient due to its small mass and the little time available before it decays. Laser cooling on 1S-2P transition has been proposed in 1988 \cite{PsLaserCooling}, but has not yet been realized because of the great challenge posed by the evanescent nature of this system \cite{PsLaserCooling2017}. 
For the 1S-2S transition of Ps, the first order Doppler shift, can be avoided using two counter propagating laser beams as done for hydrogen \cite{MPQ}. However, the second order Doppler shift cannot be avoided and results in the main systematic effect for this measurement \cite{hype15}. In order to correct for this effect the velocity of the 2S population has to be precisely determined  as  presented in this manuscript.  
\section{Description of the experiment}
Positrons from the buffer gas trap beamline at the slow positron facility at ETH Zurich \cite{Cooke2015} are extracted with 1 Hz repetition rate from the magnetic field region and implanted with 4.5 keV in 1 ns bunches of $\sim 10^4$ e$^+$ into a porous thin film $\mathrm{SiO}_2$ to form positronium. About 25\% of the positrons are converted to Ps emitted into vacuum \cite{Liszkay2008}. The output (6 mJ, 7 ns) of a pulsed dye amplifier seeded by a frequency doubled diode laser at 486 nm (the wavelength of the 1S-2S two photon transition) is synchronised with the positron beam in order to excite the atoms to the 2S state 3$\pm$0.2 mm away from the target. 
The laser is retro-reflected by a mirror to induce two photon excitation with counter propagating beams.  A significant fraction of the Ps atoms is photo-ionized in the same laser pulse by a third photon.
The surviving 2S state are further excited to the 20P state with the 736 nm output of a dye laser (3.5 mJ, 7 ns). To cover completely the first order Doppler profile of this transition, which for the given geometry of the experiment is around 150 GHz, the dye laser was set up for a bandwidth of approximately 400 GHz. Subsequently the excited atoms are detected via field  ionization (4 kV/cm) with a micro-channel-plate (MCP) after traversing 40 mm. 

By normalizing to the prompt peak due to backscattered positrons from the target, the signal area can be related to the excitation probability of a Ps atom to the 2S or the 20P level respectively \cite{PSASProceedings,WichmannPhD2018}.   

\begin{figure}[htbp]
\centering
 \includegraphics[width=0.68\textwidth]{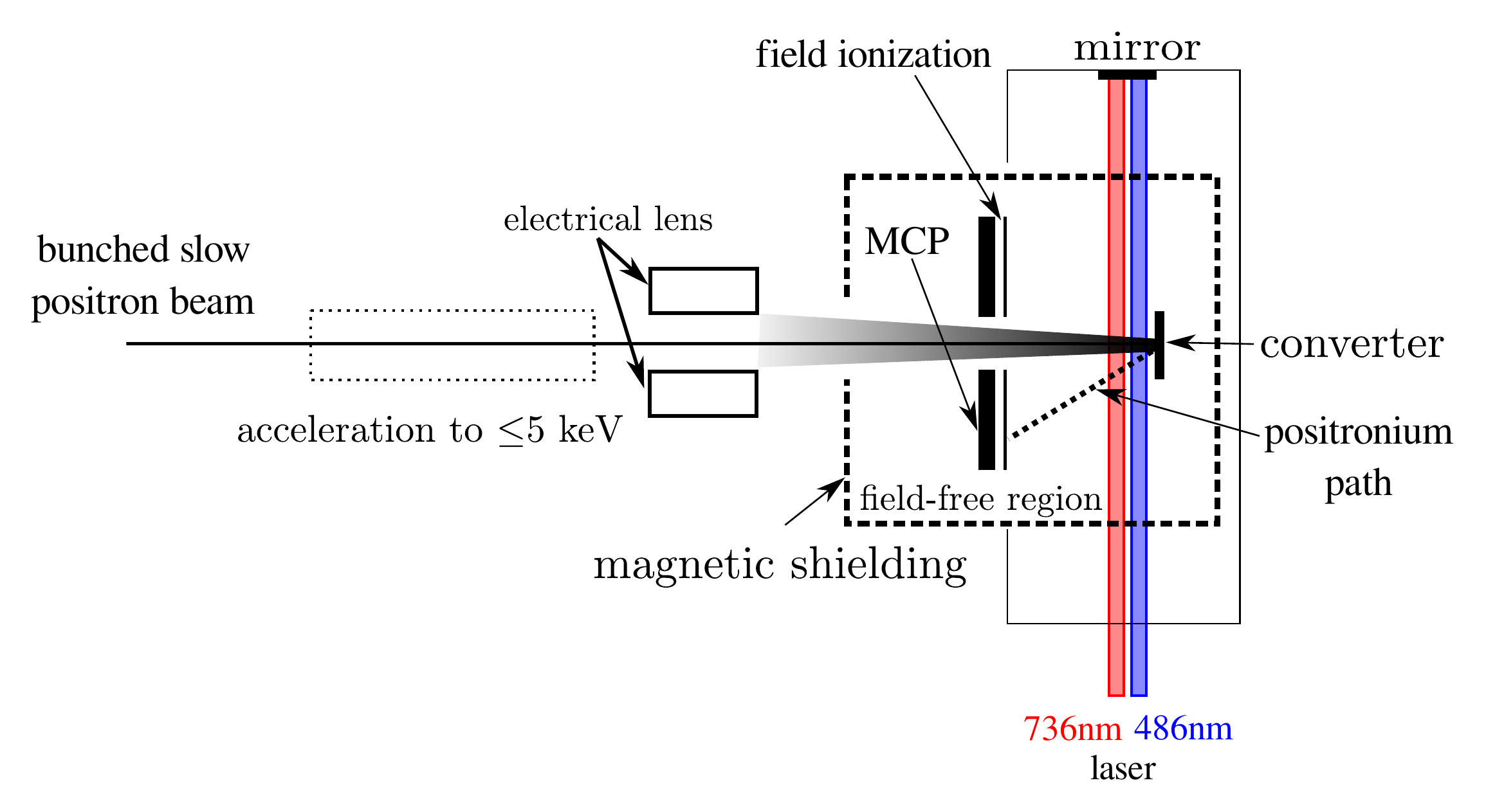}
\caption{Experimental setup and laser system}
\label{fig:expsetup}
\end{figure}

\section{Results}
\begin{figure}[htbp]
\centering
  \includegraphics[width=0.48\textwidth]{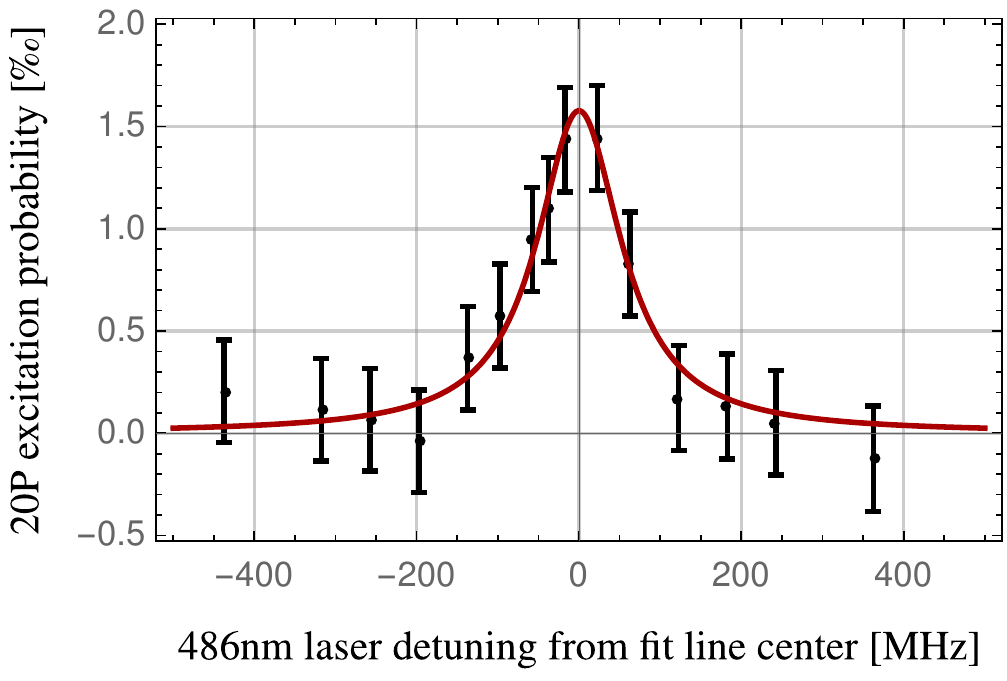}
 \caption{Resonance curve of the 2S Ps atoms detected in the MCP after excitation in the 20P and subsequent field-ionization. 
}
\label{fig:lineshape2S} 
\end{figure}

Fig. \ref{fig:lineshape2S}, shows the measured lineshape of the 1S $\to$ 2S transition detected via field ionization after the excitation of the Ps atoms from the 2S to the 20P state. 

\begin{figure}[htbp]
\centering
  \includegraphics[width=0.48\textwidth]{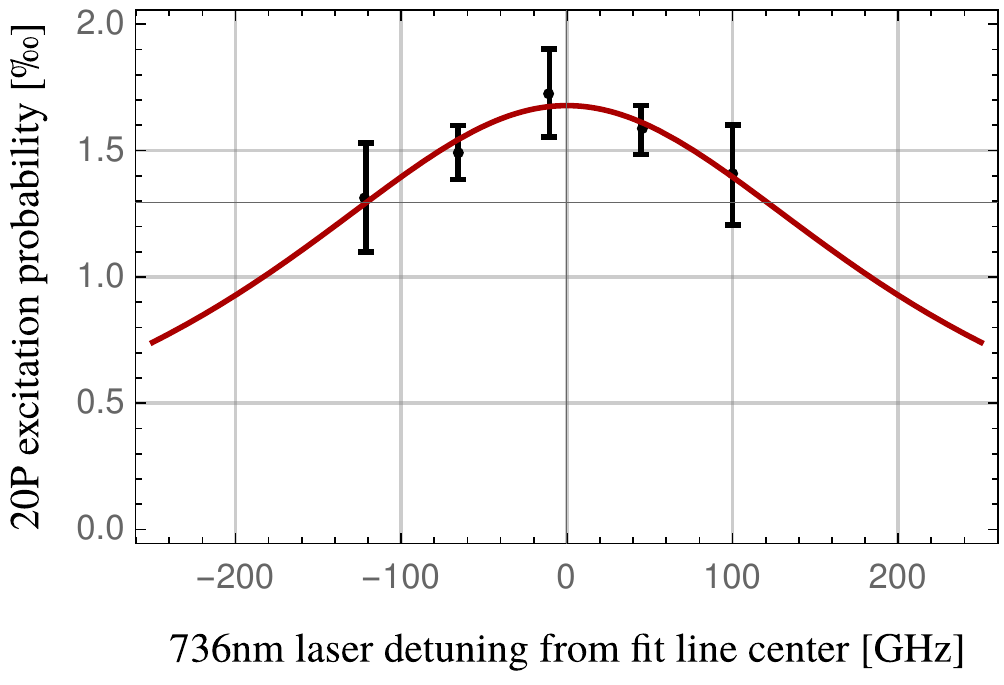}
 \caption{Resonance curve of the 2S $\to$ 20P transition, obtained by scanning the dye laser wavelength, detected by field-ionization on the MCP}
\label{fig:Scan20P}
\end{figure}

The frequency scan of the 736 nm laser is done by keeping the 2S exciting laser frequency fixed on resonance. The center of the lineshape, shown in Fig.\ref{fig:Scan20P}, is  $736.40\pm0.02$ nm in agreement with the expected wavelength of 736.39 nm. The simple fit with a Lorentzian gives a FWHM $\simeq$ 0.44 THz compatible with the laser bandwidth.

\begin{figure}[htbp]
\centering
  \includegraphics[width=0.48\textwidth]{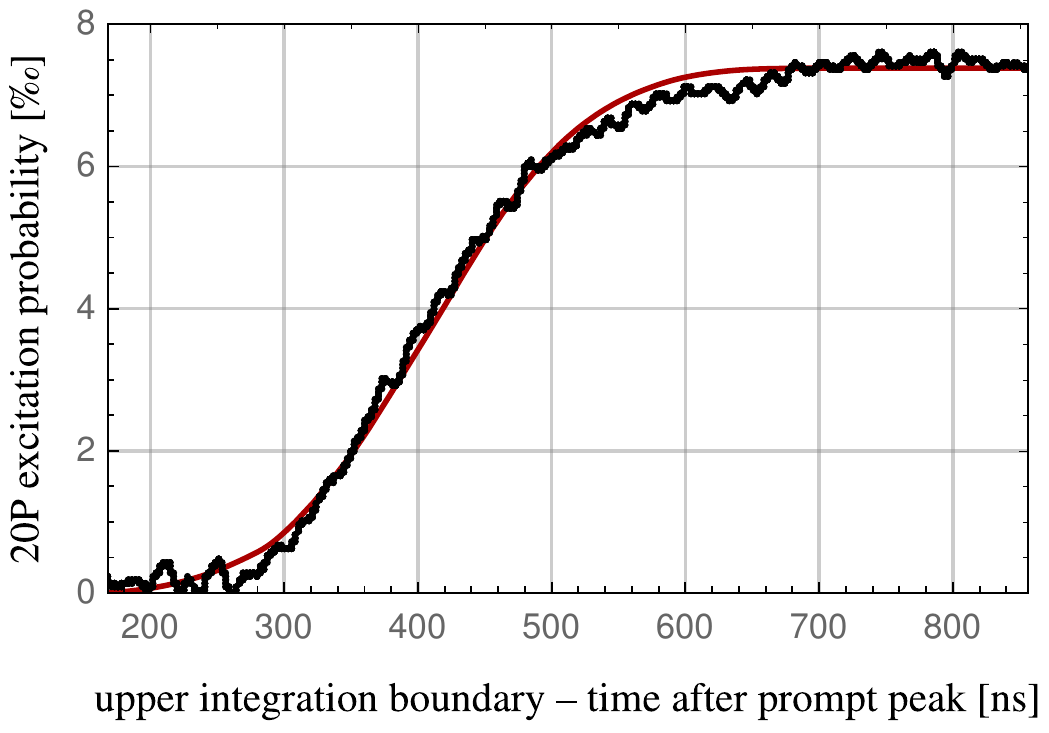}
 \caption{Measured time of flight (TOF) distribution. The red line represents the fit from the MC simulation.}
\label{fig:TOF}
\end{figure}

The measured time of flight (TOF) distribution (see Fig. \ref{fig:TOF}) is obtained by sampling the cumulative distribution function, integrating the MCP signal in the region between 150-1200 ns after the positron implantation occurs. 

\begin{figure}[htbp]
\centering
  \includegraphics[width=0.48\textwidth]{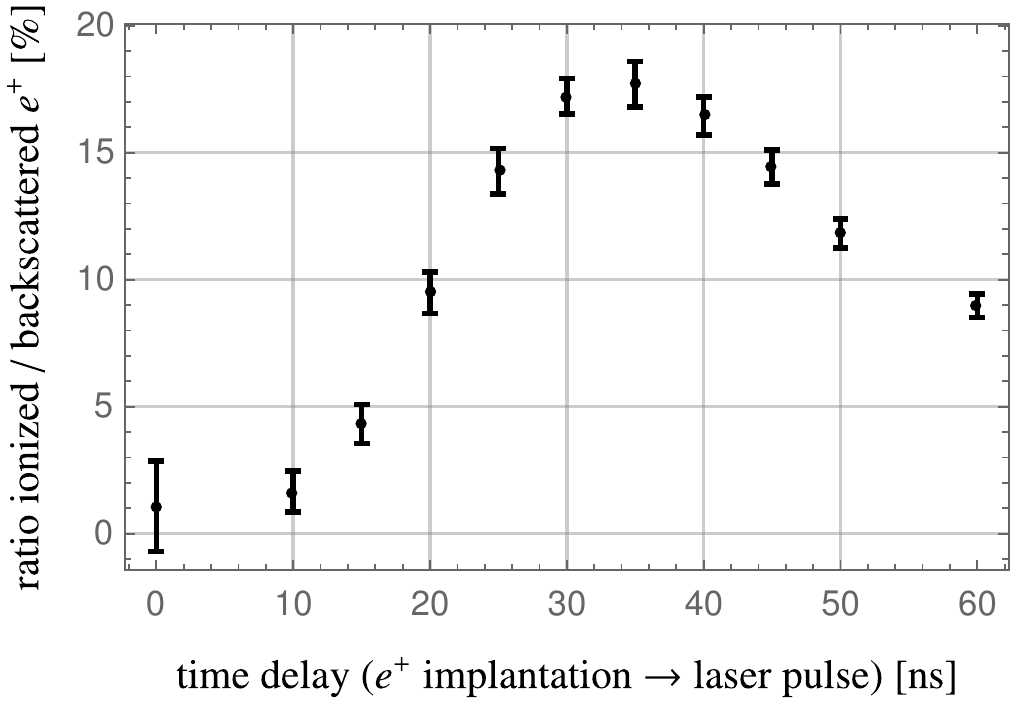}
 \caption{Excitation of the atoms in the 2S state versus the time delay between positron and laser pulses.}
\label{TimeScan}
\end{figure}

To fit the data we use a detailed Monte Carlo (MC) simulation. The interaction of positronium with the laser fields is modelled using Bloch equations that are solved numerically using the coefficients calculated for hydrogen \cite{Haas2006}, rescaled properly for positronium.
The initial distribution of positronium is a 2D Gaussian with $\sigma$=1$\pm$0.1 mm defined by the incoming positron beam as measured with a phosphor screen MCP placed behind the target that can be moved out from the beam axis. The re-emission time of Ps from the porous silica was measured by changing the delay of the laser pulse in respect to the incoming positron pulse and plotting the signal rate as shown in Fig. \ref{TimeScan}. 
These results are in agreement with previous measurements done on the same kind of samples at the given positron implantation energy of 4.5 keV \cite{CassidyDelayedPRA2010}.  
In this way, the laser delay was also optimized to obtain the largest overlap of the Ps atoms and therefore the highest yield of atoms excited in the 2S state. The measured time distributions were used as an input for the MC. The 2S excitation laser was modelled as Gaussian beam with a measured beam waist of $\sigma_{\omega_0}=2.0\pm0.2$ mm.

The porous silica films used as targets in this experiment have pore sizes of about 5 nm \cite{Liszkay2008}. This is comparable with the Ps de Broglie wavelength for sub eV Ps kinetic energies and therefore a quantum mechanical treatment is required  \cite{Brusa2008}. As it was demonstrated by different studies, the Ps is thus emitted into vacuum with an energy that is correlated with the pore size and a minimum energy of $48\pm5$ meV defined by the ground state energy in the pores \cite{CassidyPRA2010, CrivelliPRA2010}. If the positrons are implanted deep enough in the material, the Ps energy spectrum will thus be a sum of "monoenergetic" distribution reflecting the different available energy states in the material with a spread defined by the pore uniformity. This was also shown to be the case in metal-organic-frameworks \cite{oPsMOF, MillsMOF}. 
However, the positrons follow a broad implantation profile. The fraction of Ps formed near the surface will not have time to thermalize in one of the available states for Ps in the pores resulting in a fraction of the atoms emitted into vacuum with a continuuum of energies. The maximal energy depends on the formation mechanism, i.e. if Ps was formed in the bulk of the material and is then emitted into the pore by virtue of the Ps negative work function in silica (about 1 eV) or if the positron formed Ps by capturing an electron on the pore surface (mean energy of 3 eV) \cite{Nagashima1998}. The angular emission in the simulation is assumed to follow a $\cos \theta$ distribution as suggested by previous measurements \cite{CrivelliPRA2010}. This will be verified by implementing a position sensitive MCP in future measurements in order to precisely cross check the angular distribution of the Ps emitted into vacuum.

The lineshapes acquired by direct photo-ionization (PI) and delayed photo-ionization of the 2S states with a 20 mJ, 7 ns pulse at 532 nm, are fitted simultaneously together with the TOF spectra. 
To take into account the uncertainty on the parameters entering in the modelling the fit was performed with a Bayesian approach, using the Bayesian Analysis Toolkit \cite{BAT2009}. 
The results of the MC simulation are in good agreement with data and reproduce both the shift between the delayed and direct PI events and the different line broadenings arising from the AC Stark shift. As expected the events due to direct PI experienced a larger laser field and are therefore shifted towards higher frequencies. The mean energy of the atoms extracted with the fit is 60 meV which is in good agreement with the literature values measured with different techniques at the given implantation energy \cite{CrivelliPRA2010,DellerNJP2015}.  

\begin{figure}[htbp]
\centering
  \includegraphics[width=0.48\textwidth]{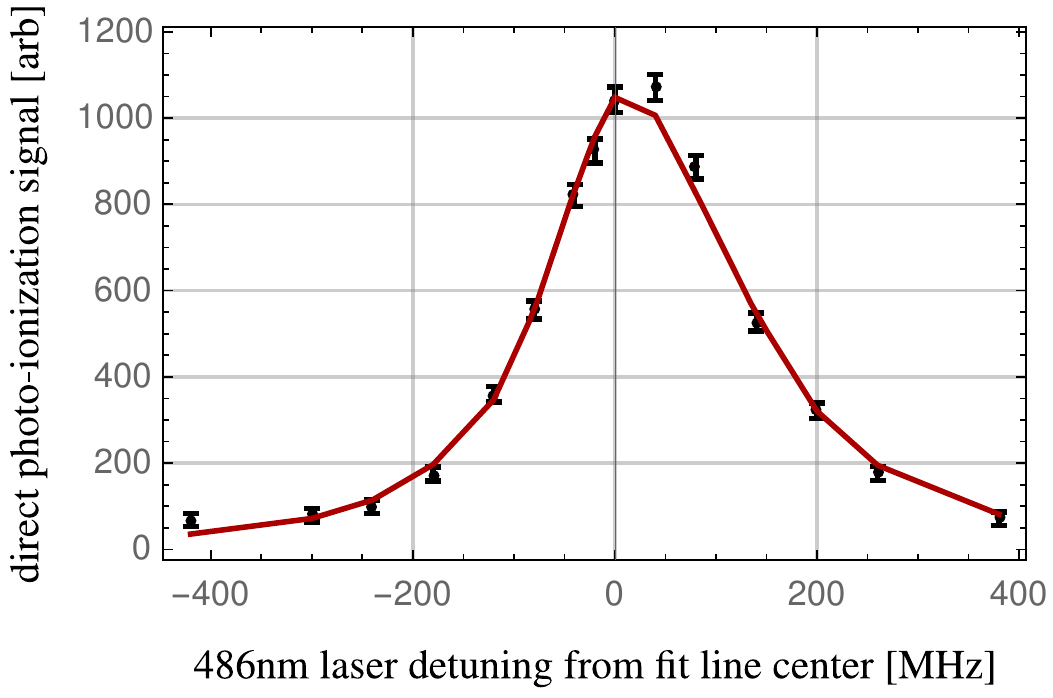}
  \includegraphics[width=0.48\textwidth]{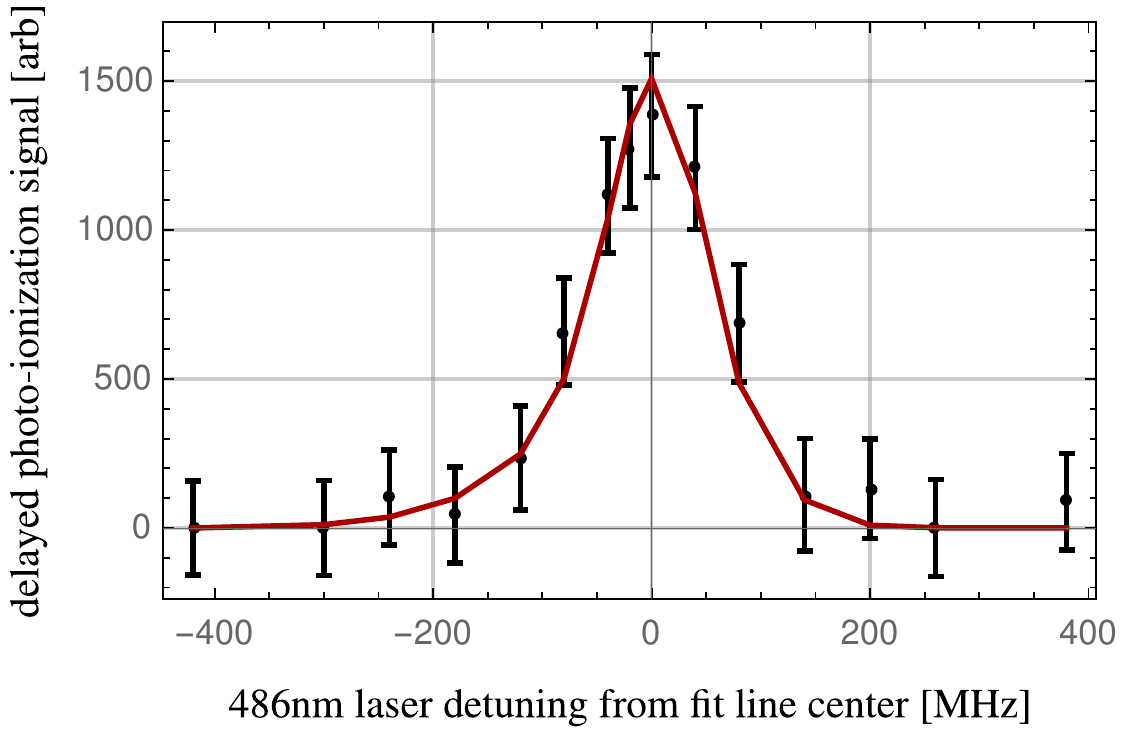}
 \caption{Lineshapes of the 2S atoms detected via direct photo-ionization (left) and delayed photo-ionization (right). The solid line is the fit of the MC to the data. }
\label{fig:MCvsData}
\end{figure}

\section{Conclusions}
In summary, we have demonstrated the excitation of Ps from the 2S to the 20P state, detected via field ionization in a MCP after $408\pm93$ ns time-of-flight. The MC simulation was validated with the data. This method allows one to reconstruct the velocity distribution of the atoms excited in the 2S state. By using this as an input for the lineshape simulation will allow to correctly take into account the main systematic uncertainty in the 1S-2S measurements of positronium arising from the second order Doppler shift.

\ack
This work was supported by the SNSF under the grant 166286 and by ETH Zurich (grant number ETH 35-14-1).  We thank  A. Antognini, B. Brown, F. Merkt, K. Kirch and A. Rubbia for their support and enlightening discussions. D. Cooke, P. Comini, L. Gerchow and A. Nguyen for their essential contributions in different stages of this project and the IPA and ETH workshops for their help.\\

\end{document}